# Goals and strategies in the global control design of the OAJ Robotic Observatory

A.Yanes-Díaz[*a], S. Rueda-Teruel[a], J.L. Antón[a], F. Rueda-Teruel[a], M. Moles[a], A.J. Cenarro[a], A. Marin-Franch[a], A. Ederoclite[a], N. Gruel[a], J. Varela[a], D. Cristobal-Hornillos[a], S. Chueca[a], M.C. Díaz-Martín[a], L. Guillen[a], R. Luis-Simoes[a], N. Maicas[a], JL. Lamadrid[a], A. Lopez-Sainz[a], J. Hernández-Fuertes[a], L. Valdivielso[a], C. Mendes de Oliveira[b], P. Penteado[b], W. Schoenell[c], A. Kanaan[d]

[a]Centro de Estudios de Física del Cosmos de Aragón Plaza San Juan 1, Planta 2 E-44001 Teruel Spain; [b]Universidade de São Paulo, IAG, Rua do Matão, 1226, Sao Paulo, 05508-900, Brazil; [c]Instituto de Astrofísica de Andalucía (CSIC), Glorieta de la Astronomía, 18008, Granada, Spain; [d]Departamento de Fisica, CFM, Universidade Federal de Santa Catarina Florianopolis, SC, 88040-900, Brazil

## ABSTRACT

There are many ways to solve the challenging problem of making a high performance robotic observatory from scratch.

The Observatorio Astrofísico de Javalambre (OAJ) is a new astronomical facility located in the Sierra de Javalambre (Teruel, Spain) whose primary role will be to conduct all-sky astronomical surveys. The OAJ control system has been designed from a global point of view including astronomical subsystems as well as infrastructures and other facilities.

Three main factors have been considered in the design of a global control system for the robotic OAJ: quality, reliability and efficiency. We propose CIA (Control Integrated Architecture) design and OEE (Overall Equipment Effectiveness) as a key performance indicator in order to improve operation processes, minimizing resources and obtaining high cost reduction whilst maintaining quality requirements.

The OAJ subsystems considered for the control integrated architecture are the following: two wide-field telescopes and their instrumentation, active optics subsystems, facilities for sky quality monitoring (seeing, extinction, sky background, sky brightness, cloud distribution, meteorological station), domes and several infrastructure facilities such as water supply, glycol water, water treatment plant, air conditioning, compressed air, LN2 plant, illumination, surveillance, access control, fire suppression, electrical generators, electrical distribution, electrical consumption, communication network, Uninterruptible Power Supply and two main control rooms, one at the OAJ and the other remotely located in Teruel, 40km from the observatory, connected through a microwave radio-link.

This paper presents the OAJ strategy in control design to achieve maximum quality efficiency for the observatory processes and operations, giving practical examples of our approach.

## 1. INTRODUCTION

The economic aspects of our engineering designs are becoming increasingly important to the extent that minimizing resources for tasks is essential whilst maintaining quality requirements. Control system design has a relevant importance in achieving these targets, not only maintaining high quality operations but also as a key for cost reduction.

[*]ayanes@cefca.es; phone +34 978 221266 fax +34 978 602334; www.cefca.es

We strongly believe that the best approach for a successful design of a new observatory is to consider it as a whole and to focus on overall efficiency basically integrated by several systems. The relationship between systems has to be optimized in order to facilitate coordination and best performance of observatory functionality as a whole.

As previously mentioned the Observatorio Astrofísico de Javalambre (OAJ[20]) is a new astronomical facility, which requires detailed engineering design work in order to achieve the required level of high performance. This document explains our general point of view of the OAJ[20] control system design.

The traditional approach of an OCS observatory control system design is to center efforts on improving operations mainly focusing on the astrophysical aspect. Our approach to the observatory control system design goes one step further because it is based on adding efficiency to the traditional approach, focusing our attention on two dimensions: What we want to obtain? (Goals) and how to obtain it? (Strategies):

Our main goal is to minimize all of the following aspects:

- **Errors:** Human, tool and system errors.
- **Resources:** Human and material resources.
- **Time:** Operation, maintenance, engineering and training time.
- **Costs:** When you reduce the previous aspects, you are indirectly reducing costs.

Our strategies to achieve these goals are the following:

- Consider the **observatory as a global entity** composed of a set of **interrelated systems**.
    - Include in this set all kinds of systems present at the observatory, not only the astronomical ones.
        - **Astronomical:** All systems exclusively related to astronomical operation.
        - **Infrastructures:** Rest of systems.
    - Give general rules and regulations for all systems regardless of the type of system.
- Provide a **global programmable tool** to control and supervise the observatory and it systems functionality. This tool must:
    - **Integrate the global architecture** of systems defined in the observatory.
    - Be **modular, flexible and scalable** in order to adapt future incorporations or modifications of systems and general structures of the observatory.
    - Give added value and functionality to **all staff profiles** or roles of people working at the observatory.
    - Close the loop providing **feedback of performance and quality.**
    - Cover a **full range of areas of functionality** to give added value to all staff profiles.
    - Integrate **all levels of implementation** including software, hardware and networking.

## 2. CIA – CONTROL INTEGRATED ARCHITECTURE

Based on previous strategies to achieve our defined main goals of minimizing errors, resources, time and costs we must develop a new programmable tool to control and manage the complete observatory. In order to perform this new design we have introduced a new concept called CIA which in our opinion summarizes the set of minimum requirements to fulfill the design of the OCS Observatory Control System in order to add and obtain high performance in the observatory's functionality. In other words the CIA concept extends the functionality of classical OCS to achieve the main goals by adding the following requirement criteria.

## 2.1 Control

Control is the first concept in CIA **Control** Integrated Architecture, but in this case it does not have the classical meaning of control loop applied to a system, in this case the loop is applied to the users of the tool. In fact it is related to closing the loop, giving feedback of performance and quality to the staff working at the observatory; on the one hand for taking short-term decisions to improve management and operations and on the other for taking long-term decisions to improve maintenance and engineering for the whole system.

The main idea of this point is that to reach our goals in fact what we really need is to improve our system, minimizing errors, resources, time and costs. Therefore the first thing we have to do is measure, taking into account the condensed phrase: "If you cannot measure it you cannot improve it" which comes from "If you cannot measure it you cannot control it"[1] and "If you cannot measure it, then it is not science"[2].

In conclusion if we want to improve performance, we have to look for a way to measure it. It is exactly the same for quality. If we want to improve quality, then we have to implement something to measure it. As a result two important modules have to be implemented as part of the OCS:

- **KPI:** Key Performance Indicators. Feedback of actions and processes to maintain high performance throughout the entire life cycle of the observatory and subsystems.

- **QI:** Quality Indicators and alarms to obtain in real time or as historical data for a certain period of time a complete overview and re-evaluation of the specification of actions and processes performed at the observatory.

In the case of quality this is commonly done mainly for astronomical operations software. This is not the case of performance in astronomy but it is quite easy to find several references in other fields of knowledge such as industrial manufacturing engineering as we will see in the following section.

Not only is it necessary to have feedback on performance and quality, but it is also important for procedure control to be implemented as part of the OCS.

- **Procedures:** Management of guiding actions to help and perform defined protocols at the observatory in order to maintain quality standards.

At this point CIA will also provide classical tools to perform global system analysis where users will have the capability to perform global and local system diagnosis in real time for short-term decisions and also a historical analysis for long-term decisions.

- **Global System Analysis Tools:** General classical tools such as System Reports, Summarized Alarms, Historical Alarms, Real Time Trends, Historical Trends, Pareto diagrams.

The process of improvement is the result of an iterative sequence of reengineering decisions coming from objective data, but the important thing is not to know what we have to do, but to know what we have to do first because our money and time is limited. For example doing a downtime analysis KPI has a lot to do in this sense with helping us to understand system behavior using objective data and not based on impressions or other kinds of subjective opinions.

## 2.2 Integrated

Integrated is the second concept in CIA Control **Integrated** Architecture and is related to the scope of global system design; aspects that need to be studied in detail, such us integration, standardization, modularity, scalability and flexibility.

- **Integration:** Two main concepts are required at this level.
    - **Systems integration:** All systems must be included as a global concept, this is our main premise because as we have said before all systems are interrelated and if something is missing then it is not a real model of the observatory. At this point it is necessary to remember that our aim is to improve the observatory's performance by reengineering. To do this the analysis of systems interrelations, behavior and responses is absolutely essential. Nature is not linear and often two different, apparently unrelated systems have a strange relationship. There is however, a way of discovering this by performing a root-cause analysis but it is absolutely essential to include initial data for both systems.

- **Layers integration:** Basically the idea is to integrate all levels of implementation to control the observatory and its systems from low-level (physical level) to high-level (application level). This is the hardest concept because obviously these layers are easily separable, but the meaning we want to transmit is to adapt the layers even before design and construction. This is because an observatory is a life system in that it is changing every day, so preparing a good infrastructure in advance in order to adapt easily to the demanding changes is vital. If we consider these layers as horizontal layers, at any time we have to be able to vertically include or remove features and functionalities in the global system. This is to say that the hardware layer is almost a software layer because the basic infrastructure, rules and regulations are defined from scratch to obtain flexibility for adapting to new needs.
    - **Software layer:** High-level programmable layer for controlling all systems at the observatory.
    - **Network layer:** Communications layer interconnecting systems with hardware layer and software layer.
    - **Hardware layer:** Low-level programmable layer for controlling each system at the observatory.

The OAJ control System has been designed with EtherCAT[12] technology, a deterministic fieldbus network connecting all of the buildings at the observatory via multi-mode optic glass fiber 50/125 µm ring. EtherCAT[12] enables control concepts that could not be realized with classic fieldbus systems. The high bandwidth enables status information to be transferred with each data item. The bus system is not the "bottleneck" of the control concept. Distributed I/Os are recorded faster than is possible with most local I/O interfaces. A second fiber optic network is also installed at the OAJ in this case with redundant star topology for Ethernet communication between buildings. In both cases the selected fiber optic immunizes the network layer infrastructure against occasional rays.

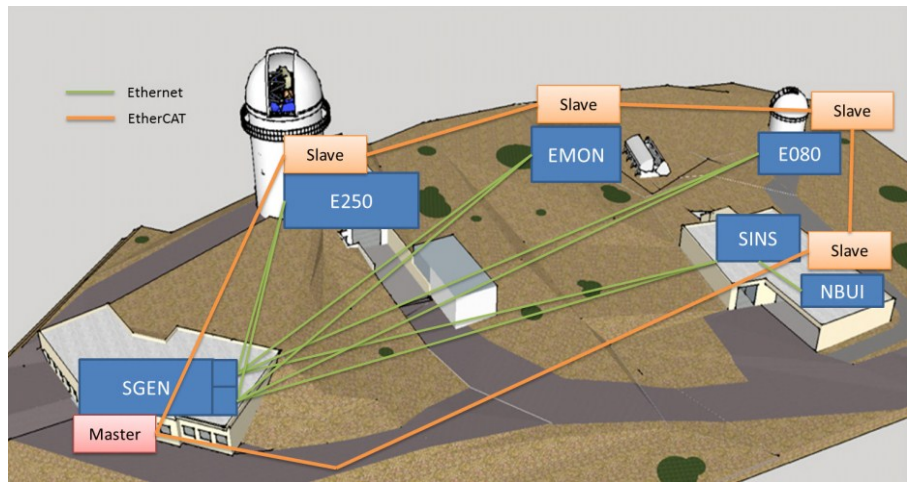

Figure 1

Figure 1 shows a representation of the main EtherCAT and Ethernet networks; secondary networks are not represented in order to facilitate interpretation.

On each building there is a Beckhoff[11] EtherCAT slave node connected to the main EtherCAT network. On the one hand this node is communicating data with a secondary EtherCAT network interchanging all signals coming from all different PLC systems present in each building, on the other hand this node is communicating data with the main Beckhoff[11] PLC which is the master of the main EtherCAT network and is in charge of concentrating all signals at the observatory in order to interchange data with EPICS[8] SCADA.

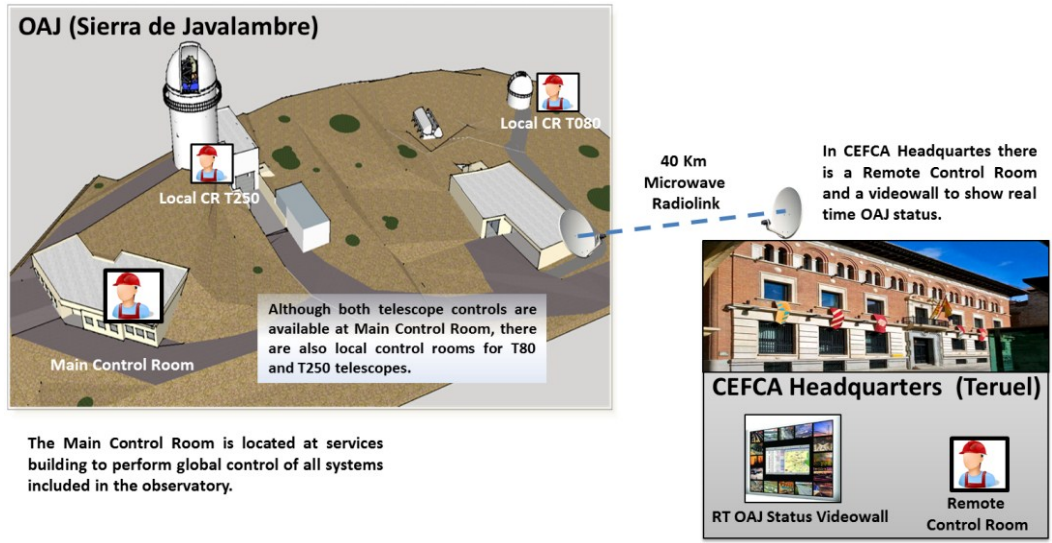

Figure 2

Figure 2 shows control rooms' deployment. There are four control rooms in total, three located at the observatory, one is the main control room, then there are two local control rooms, one for each telescope and the last one located in CEFCA Headquarters, from this remote control room is possible to perform the same tasks than in the main one.

- **Standardization:** Standards are a powerful tool for the observatory because there are several different kinds of systems of all sizes; therefore standardization is a must if we want to manage everything with efficiency because it represents improvement in universal technical communication and mutual understanding.
    - o  Our first step is to standardize terminology, this is necessary in order to easily sort, filter and find the amount of systems signals deployed at the observatory.

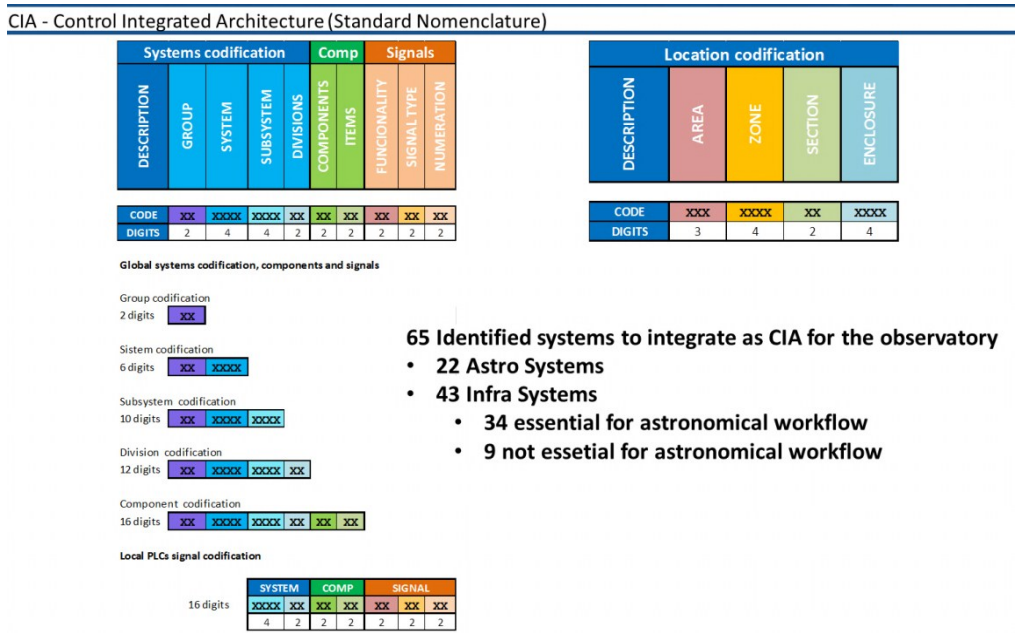

Figure 3

Figure 3 shows defined codification of nomenclature to use at all levels of CIA OAJ control system.

- o The second step is to homogenize software as much as possible: servers, clients, functionalities, screens, navigation, databases, alarms, trends, reports, etc. This is also positive as it reduces training costs and learning time because all of the staff are using the same tool with similar procedures to operate, manage, maintain and analyze different systems.

- o The third step for our tool is to select standards for different kinds of devices, brands, protocols or other entities present at the OCS Standardization, minimizing spare parts and development time because they can be shared by different systems. The selection for OAJ control system design is shown in Table 1

Table 1. Actual selection for OAJ control system design.

| ITEM | DESCRIPTION |
| --- | --- |
| Operative System | Linux CentOS[3] under VMware ESXI[4] |
| Networking | Ethernet TCP/IP (CISCO[5], FO double star topology, Main Core: 4507R, Secondary Switches: 3750X) |
| Radio link | SAGEMCOM[6] Microwave Radiolink SLF-H (700Mbs, 23GHz and 6GHz bands) |
| Servers and PCs | Dell[7] PowerEdge R510 Servers and thin clients |
| Languages | C++, Java, Python |
| Command line and scripting module | Chimera[22] (Own design developed by Brazilian collaborators) |
| SCADA | EPICS[8] open source SCADA |
| HMI | LabView[9] (Local and temporal HMI until EPICS project has been finished) |
| PLC software | TwinCAT[10] embedded IEC 61131-3 |
| PLC hardware | Beckhoff[11] (C51xx for main PLC data concentrator and local control with CX50x0 and CX10x0 series) |
| Global fieldbus | EtherCAT[12] multi-mode optic fiber ring through all buildings at OAJ |
| Local fieldbus | EtherCAT[12] cooper line, Profibus-DP[13], CANbus[14] |
| Process control protocol | OPC (Matrikon[15] and Kepware[16] only for special systems through SNMP[17]) ADS ( Own design developed by CEFCA) |
| Motors | Etel[18] |
| Encoders | Heidenhain[19] |

- o Finally system function requirements and characteristics are normalized and unambiguous testing methods are established for all systems.
- **Adaptability:** An astronomical observatory is a facility which continuously grows and changes in order to adapt to new project specifications. Therefore, the OCS has to be prepared to make changes easily, and the selection and implementation of the initial technology is a critical factor because it will describe the future working lines. Therefore a modular, scalable and flexible tool has to be implemented.

- o **Modular:** The structure of the OCS has to be made of different separable elements with the capability to work independently as standalone functionality. In addition, at a physical level, the modular applicable technology results in a reduction of stocks of spare parts and therefore in lower storage costs.

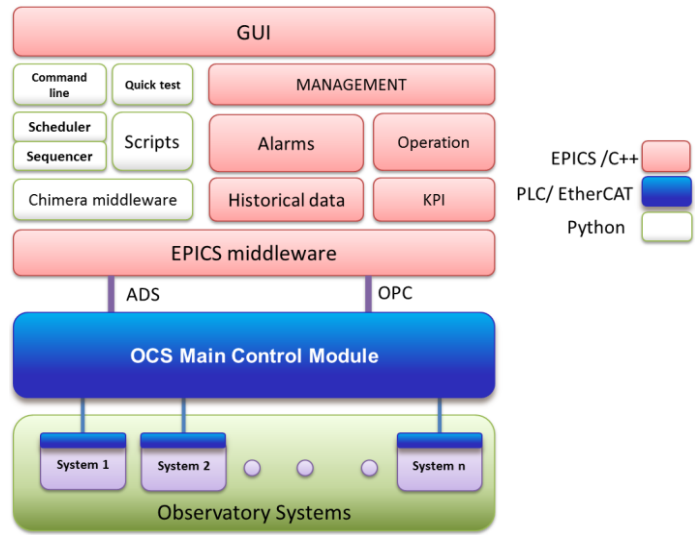

Figure 4

Figure 4 shows defined modules for OAJ control system where main software is based on EPICS[8] SCADA with C++, some software modules are being developed in python in order to offer tools in a wide extended language for astronomers. Chimera[22], Scheduler[23] and Sequencer[23] are included in these pythons modules.

- o **Scalable:** OCS must be easily customized to the actual requirements and scaled up later if necessary.
- o **Flexible:** Flexible engineering of the I/Os when a project is continuously changing implies highly reliable planning. It is common for all systems to be reengineered so that they are dynamic and resilient. Being able to change control techniques easily results in significant benefits in terms of performance and life-cycle costs.

## 2.3 Architecture

Architecture is the third concept in CIA Control Integrated **Architecture** and is related to building a tool which can give functionality in all areas covering the wide range of staff profiles requirements at the observatory.

- Give added value and functionality to **all staff profiles** working at the observatory.
    - o Managers
    - o Scientist
    - o Astronomers
    - o Operators
    - o Engineers
    - o Maintenance
- Cover a **full range of areas of functionality** to give added value to all profiles:
    - o **Management:** Global administration of all systems at the observatory.

- o **Operations:** Actions and processes performed at the observatory by a certain group of systems such as astronomical systems, in this case called astronomical operations.
- o **Engineering:** Analysis, design, development and implementation of all systems at the observatory.
- o **Maintenance:** Supporting tasks to keep all systems at the observatory in good condition.

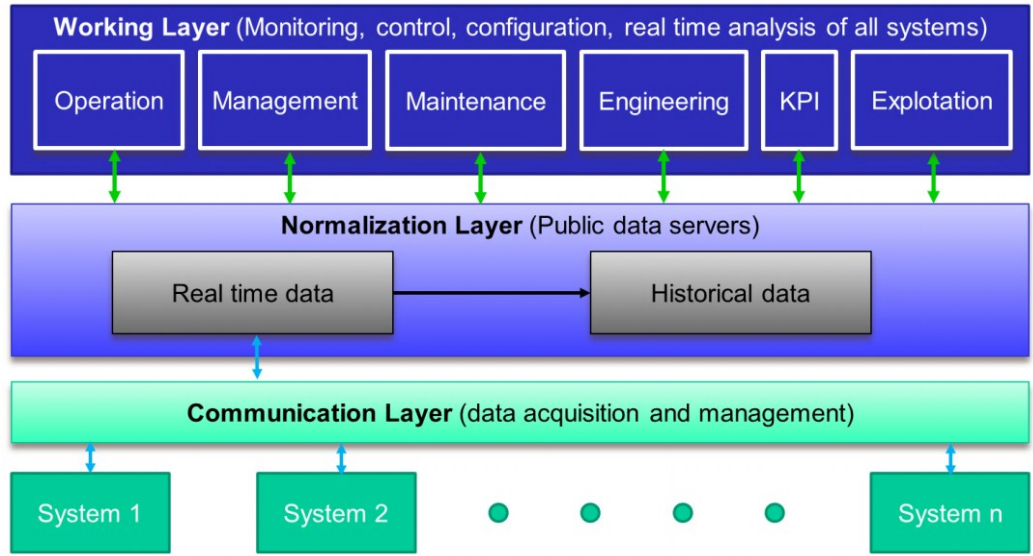

Figure 5

Figure 5 shows a diagram of CIA architecture defined for OAJ control system

## 3. OEE - OVERALL EQUIPMENT EFFECTIVENESS

As we mentioned at the beginning of this paper economic aspects in astronomical observatories are becoming increasingly important to the extent that minimizing resources for tasks is essential whilst maintaining quality requirements.

The question is "How can an observatory optimize the performance of their existing facilities?". Our answer to this question is to use Overall Equipment Effectiveness (OEE[21]) as a KPI (Key Performance Indicator) for the observatory adapting it to the special and detailed needs of an astrophysical observatory.

OEE[21] is a well-known tool, developed by Seiichi Nakajima[21] in 1960's for the automation manufacturing industry and it is currently the most implemented in the automation industry to improve performance. Initially it appears that the two fields are completely different and as a matter of fact they are, but there is something that both (observatories and the manufacturing industry) have in common; both have to minimize downtimes, OEE is directly related to study downtimes and other wasting time analysis.

We truly believe that OEE is an effective tool to benchmark, analyze, and improve observatory processes. The OEE tool provides the ability to measure systems for performance improvements. OEE not only measures these inefficiencies but groups them into three categories helping to analyze the system and giving a better understanding of the observatory process.

The author would like to comment on the ease of adapting this tool to an astrophysical observatory, particularly so for survey projects. Therefore, taking into account that the OAJ's[20] main goal is to perform a survey project, in the OCS design we have implemented OEE as an essential KPI for the OAJ control system.

The OEE is an objective measured value of performance, is not an absolute index, is relative to the own system, in fact the good thing is that you can compare the OEE of two different telescopes in order to evaluate the difference of their performance, or compare the OEE of the same telescope but in two different periods of time. Although the OEE is calculated in real time to know actual OEE for example of this night, you can also obtain the historical OEE of a certain period of time, in other words the last night OEE, or last week, or last year.

The OEE will encompass the whole survey flow. It is a tool which gives information to all staff profiles because everybody has a direct impact on some part of the process. Figure 6 shows an example of real time OEE indicators.

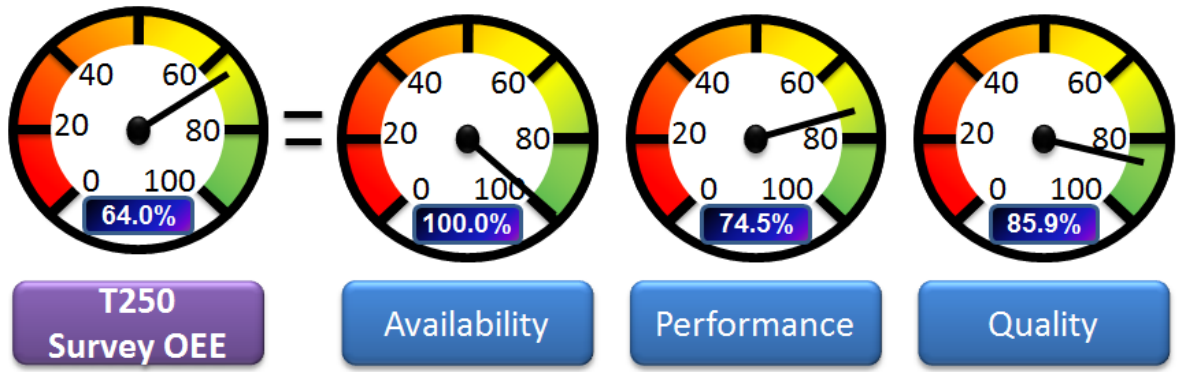

Figure 6

The OEE is very easy to understand because is a percentage going from 0% to 100%, when OEE is 0% means that something is going really bad, when is 100% it means that everything is going absolutely perfect. In fact obtain an OEE value of 100% is not a real situation that value is an ideal situation otherwise OEE implementation is not well done. A more realistic situation with good performance would be something like an OEE moving around 80%. Do not worry if you implement OEE at your observatory and you find out that you have an OEE around 30% almost every day. In fact this is a good new because it means that you have a lot of work to do in order to improve your performance.

The design of the OAJ control system foresee three OEE calculations, one for the T250 telescope, another for the T80 telescope and the last one a global OEE for the whole observatory.

### 3.1 OEE Implementation

In order to improve the performance of equipment, it is necessary to determine its current state through the analysis of its rates. Nakajima[21] used only three basic indicators to summarize the main factors that cause time loss when using equipment.

$$OEE(\%) = Availability(\%) \times Performance(\%) \times Quality(\%)$$

Equation 1

Equation 1 shows the three contributions to the OEE. The important thing is to understand what is happening in your system when you obtain a value of OEE. For example, imagine you obtain a value of OEE=8.1% then the second step is to analyze availability, performance and quality independently; because there are many possibilities:

- A=10%,P=90%,Q=90% in this case OEE is mainly affected by availability
- A=90%,P=10%,Q=90% in this case OEE is mainly affected by performance
- A=90%,P=90%,Q=10% in this case OEE is mainly affected by quality
- Many other combinations could be possible and you need to understand in detail what is happening.

To continue with the explanation of the OEE adaptation to the observatory is necessary to define working modes for all systems at the observatory, this is something coming from CIA concept where standardizes general modes for all systems integrated or a group of them. Figure 7 shows general defined modes and states for CIA systems at the OAJ.

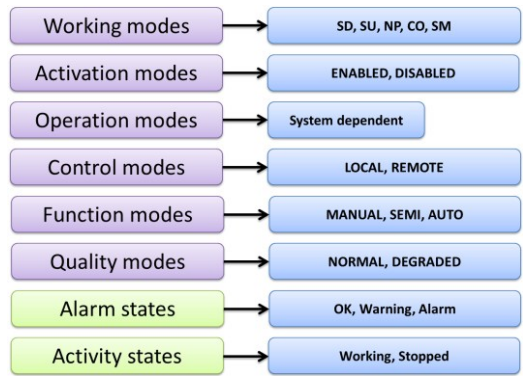

Figure 7

First general mode is working mode and this is absolutely necessary for OEE understanding. SD=Scheduled Down is the working mode where the system is not foresee to be working. SU=Start Up and CO=Change Over are only transient states and NP=Normal process the system is foresee to be working (maybe is not working because bad weather, but the mode will be NP and time is counting to NP) SM= Sanitation and Maintenance the system is foresee to be working but only for maintenance or engineering task. For example in the case of OAJ T250 telescope NP is the mode where is foresee to be doing the survey and during the day working mode could be SM or SD. Figure 8 shows possible working modes flow for systems present at the observatory.

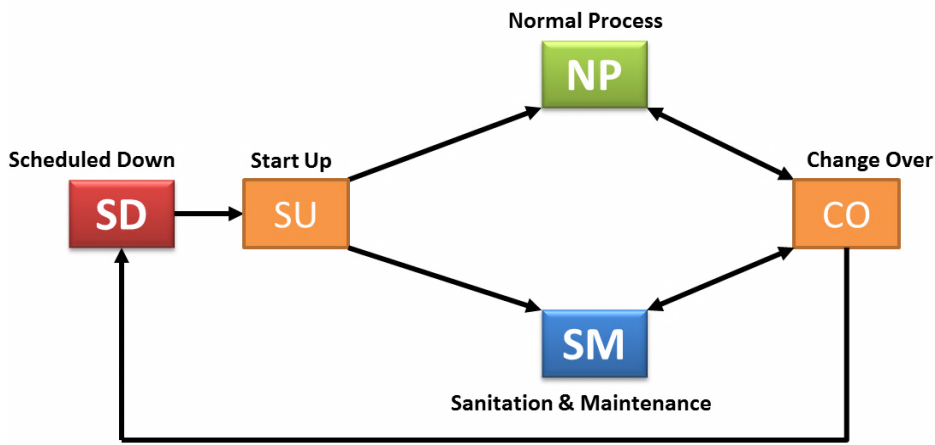

Figure 8

Figure 9, 10 and 11 shows our proposal of OEE parameters in order to adapt to observatory survey functionality.

In figure 9 is shown Availability parameter, and is represented how we lose time from availability due to contributions such as weather, downtimes, breakdowns, configuration, tuning, startups or change overs working modes. At first from total time we only consider scheduled time which is the time of SU+NP+CO working modes, in fact is the total theoretical time available to work, in other words for us will be our astronomical night.

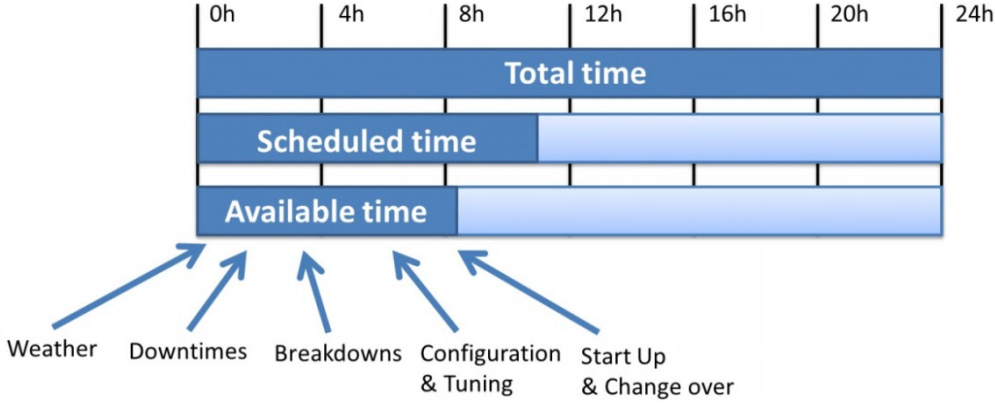

Figure 9

In figure 10 is shown Performance parameter, and the contribution to lose time is due to a different type of factors such as operation, micro_downtimes, bad survey strategy and technical failures.

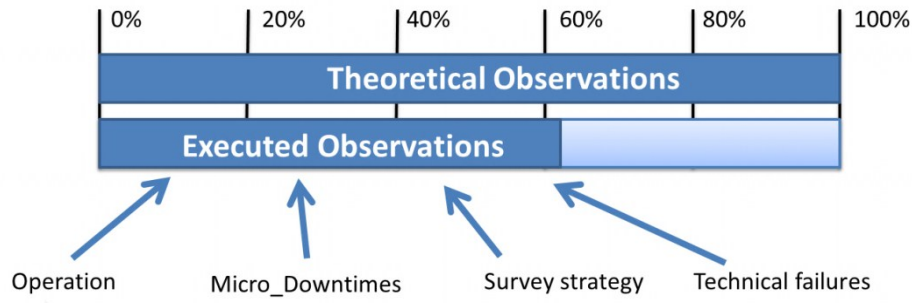

Figure 10

In figure 11 is shown Quality parameter, and the contribution to lose OEE is due to a quality factors such us seeing, extinction or other effects such as bad focusing.

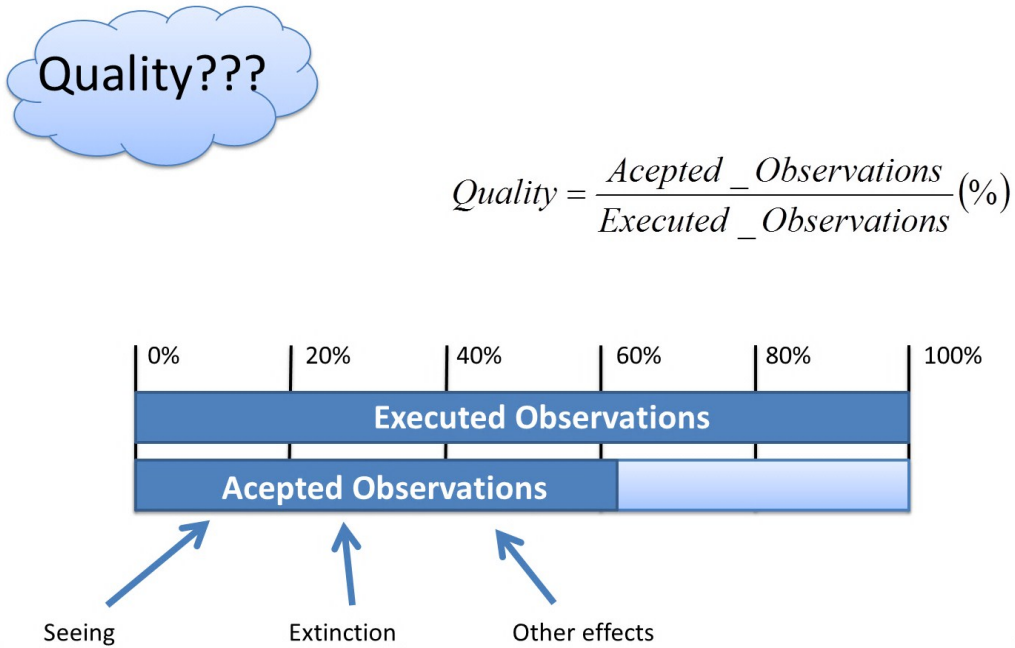

$$Quality = \frac{Acepted\_Observations}{Executed\_Observations}(\%)$$

Figure 11

## 4. CONCLUSIONS

This paper has presented two important concepts in the design of the Observatory Control System in order to provide the observatory the capability to achieve high performance in processes execution. First concept is CIA Control Integrated Architecture and it is a set of global requirements to fulfill the OCS. Second concept is the OEE and our proposal is to include it as a key performance indicator for astrophysical observatories and specially indicated for surveys. We are convinced that CIA specifications and OEE will add good value to OCS Observatories Control Systems in order to be better prepared for present and future needs.

The development and implementation of OAJ control system has 6 different overlapping phases: Phase 0: Observatory construction, Phase 1: Independent and Local control systems, Phase 2: Integrated Architecture, Phase 3: Global Systems Management, Phase 4: Global Systems Analysis & Diagnosis, Phase 5: Systems Performance Improvement.

At the time of writing this paper the development status of the OAJ control system is the following:

We are working on Phases 0, 1 and 2. Design, implementation and verification of the infrastructure network layer have been done. Design of software and hardware layers are finished and now we are implementing and integrating with CIA requirements first group of systems (among which is included first telescope T80) at OAJ control system.

Following steps are to continue developing and implementing OAJ control system with EPICS middleware, setting up historical servers, alarms, sequencer[23], scheduler[23] integrating Chimera[22] middleware and then start with Phase 3.

# ACKNOWLEDGMENTS

The OAJ is funded by the Fondo de Inversiones de Teruel, supported by both the Government of Spain (50%) and the regional Government of Aragón (50%). This work has been partially funded by the Spanish Ministerio de Ciencia e Innovación through the PNAYA, under grants AYA2006-14056 and through the ICTS 2009-14, and the Fundación Agencia Aragonesa para la Investigación y Desarrollo (ARAID).

This project is also partially funded by Brazilian institutions CAPES, CNPq, FAPESP and INCT-A.

Setting up an astrophysical observatory is a hard task which demands a huge effort from all of the staff. The author would like to acknowledge the dedicated work of the entire CEFCA team over the last two years. The remarkable success of the current state of the observatory is the result of a true team effort. This visible success gives us added energy to continue with the construction and development of the observatory.

The author would also like to acknowledge the extraordinary quality of the work that the AMOS team are doing. I would particularly like to mention Olivier Pirnay, Eric Gabriel, Bruno Rongier, Jean-Marc Tortolani and Gregory Lousberg from AMOS Ltd in Liège, Belgium.

Finally, the author would like to thank Miguel Nuñez and Josefina Rosich from IAC Canary Islands, Spain for their valuable help. Their kind attention has been much appreciated.